\documentclass[twoside]{IEEEtran}
\usepackage{graphicx}
\usepackage[T1]{fontenc}
\usepackage{graphicx}
\usepackage{adjustbox}
\usepackage{listings}
\usepackage{url}
\usepackage{booktabs}
\usepackage{xr}
\usepackage{longtable}
\usepackage{algorithmic}
\usepackage{amssymb}
\usepackage{multirow}
\usepackage{glossaries}
\usepackage{setspace}
\usepackage{float}
\usepackage{caption}
\usepackage{soul}
\usepackage{subfig}
\usepackage{xcolor}
\usepackage{enumitem}

\makeglossaries

\newacronym{WWW}{WWW}{World Wide Web}
\newacronym{MCC}{MCC}{Mobile Cloud Computing}
\newacronym{MCE}{MCE}{Mobile Cloud Environment}
\newacronym{QoS}{QoS}{Quality of Service}
\newacronym{QoE}{QoE}{Quality of Experience}
\newacronym{OaaS}{OaaS}{Offloading as a Service}
\newacronym{AOSP}{AOSP}{Android Open Source Project}
\newacronym{DVM}{DVM}{Dalvik Virtual Machine}
\newacronym{TDP}{TDP}{Thermal Design Power}
\newacronym{GCC}{GCC}{GNU Compiler Collection}
\newacronym{ALVM}{ALVM}{Application Level Virtual Machine}
\newacronym{ALVMs}{ALVM}{Application Level Virtual Machines}
\newacronym{ELF}{ELF}{Executable and Linking Format}
\newacronym{VMM}{VMM}{Virtual Machine Monitors}
\newacronym{DSM}{DSM}{Distributed Shared Memory}
\newacronym{SOA}{SOA}{Service Oriented Architecture}
\newacronym{JIT}{JIT}{Just in Time}
\newacronym{ART}{ART}{Android Runtime Environment}
\newacronym{AHOT}{AHOT}{Ahead of Time}
\newacronym{JVM}{JVM}{Java Virtual Machine}
\newacronym{CLR}{CLR}{Common Language Runtime}
\newacronym{VM}{VM}{Virtual Machine}
\newacronym{IDE}{IDE}{Integrated Development Environment}
\newacronym{OSP}{OSP}{Offloading Service Provider}
\newacronym{PMCO}{PMCO}{Process Migration based Computational Offloading}
\newacronym{WiFi}{WiFi}{Wireless Fidelity}
\newacronym{GPRS}{GPRS}{General Packet Radio Service}
\newacronym{LTE}{LTE}{Long Term Evolution}
\newacronym{WAN}{WAN}{Wide Area Network}
\newacronym{TCP}{TCP}{Transmission Control Protocol}
\newacronym{UDP}{UDP}{User Datagram Protocol}
\newacronym{HTTP}{HTTP}{Hyper Text Transfer Protocol}
\ifCLASSINFOpdf
\else
\fi
\begin{document}
\title{Process migration-based computational offloading framework for IoT-supported mobile edge/cloud computing}

\author{Abdullah~Yousafzai,
        Ibrar~Yaqoob,~\IEEEmembership{Senior Member,~IEEE,}
        Muhammad~Imran, 
        Abdullah~Gani,~\IEEEmembership{Senior Member,~IEEE,}
        and~Rafidah~Md~Noor 
        
        % <-this % stops a space
\thanks{Abdullah Yousafzai is with the Department
of Computer Systems, University of Malaya, Kuala Lumpur, 50603, Malaysia. (e-mail: abdullahyousafzai@siswa.um.edu.my).}% <-this % stops a space
\thanks{Ibrar Yaqoob is with the Department of Computer Science and Engineering,
Kyung Hee University, Yongin-si 17104, South Korea.}% <-this
\thanks{Muhammad Imran is with the College of Computer and Information Sciences, King Saud University, Saudi Arabia.}% <-this% stops a space
\thanks{Abdullah Gani and Rafidah Md Noor are with the Department
of Computer Systems, University of Malaya, Kuala Lumpur, 50603, Malaysia.}% <-this
}


\markboth{IEEE Internet of Things Journal}%
{}
%Yousafzai \MakeLowercase{\textit{et al.}}: A process migration based computational offloading framework for mobile edge/cloud computing



\maketitle

\begin{abstract}

Mobile devices have become an indispensable component of Internet of Things (IoT). However, these devices have resource constraints in processing capabilities, battery power, and storage space, thus hindering the execution of computation-intensive applications that often require broad bandwidth, stringent response time, long battery life, and heavy computing power. Mobile cloud computing and mobile edge computing (MEC) are emerging technologies that can meet the aforementioned requirements using offloading algorithms. In this paper, we analyze the effect of platform-dependent native applications on computational offloading in edge networks and propose a lightweight process migration-based computational offloading framework. The proposed framework does not require application binaries at edge servers and thus seamlessly migrates native applications. The proposed framework is evaluated using an experimental testbed. Numerical results reveal that the proposed framework saves almost 44\% of the execution time and 84\% of the energy consumption. Hence, the proposed framework shows profound potential for resource-intensive IoT application processing in MEC.

\end{abstract}


% \fi
\IEEEpeerreviewmaketitle

\section{Introduction}
\IEEEPARstart{T}{he} remarkable proliferation of resource-intensive Internet of Things (IoT) applications, such as face recognition, ultrahigh-definition video, voice semantic analysis, interactive gaming, and augmented reality, have gained immense popularity in recent years \cite{wang2019edge23, hassan2018role}. These IoT applications require intensive computation capabilities. However, smart mobile devices that support IoT have limited computational resources and battery constraints compared with desktop computers. Consequently, these limitations pose considerable challenges for future mobile platform development \cite{govindan2019tcp}. Computational offloading is an appealing software-level solution that helps mitigate the problem of resource-constrained mobile devices by enabling the remote execution of computation-intensive IoT applications \cite{8684800,yu2017survey}. These applications are offloaded to the edge of the user network, which shortens the execution time compared with a typical cloud computing environment \cite{chiang2016fog, abedin2018resource}. We use the term "mobile cloud environments" (MCE) to generalize mobile cloud and mobile edge computing (MCC and MEC, respectively).

Conventionally, a computational offloading architecture is composed of a client and a server subsystem. The former is configured and executed on the mobile/IoT device, whereas the latter is available either on the network edge \cite{8016573} or the cloud provider \cite{Chun:CloneCloud,cuervo2010maui}. The client subsystem performs three major tasks to optimize the net system utility: i) observes and estimates the network performance metrics for the mobile device; ii) monitors, estimates, and analyzes the resource requirements of mobile applications in terms of CPU time on the mobile device and the cloud server; and iii) generates task-migration-related decisions \cite{OaaS}. The server subsystem is generally supported by the cloud computing business model, which revolutionizes the business life cycle by reducing the capital investment in infrastructure while maintaining additional focus on business services and strategies \cite{nist_def}. Given these prominent features, cloud computing systems have expanded into the business model of MCC \cite{ahmed2019process}, which is further extended to MEC \cite{khan2019edge, 8114564}.

Code migration is one of the most common computational offloading mechanisms, which migrates intermediate-level instructions between a mobile device and a server \cite{yousafzai2016computational}. These instructions must be executed on the same type of application-level virtual machines (ALVMs) of the mobile device and server  \cite{yousafzai2016computational}. Literature also reveals offloading mechanisms that consider thread-state migration or thread-state synchronization. Code migration and thread-state synchronization highly depend on ALVMs. This dependency invalidates the offloading mechanisms for native application binaries. For example, Google has offered android runtime environment (ART), which introduces ahead of time (AHOT) compilation to platform-dependent native machine instructions to replace the just-in-time-based Dalvik virtual machine (DVM). ART is beneficial in terms of execution time and battery consumption and uses AHOT compilation to transform device-independent DEX code into device-specific machine binaries  \cite{yousafzai2016computational}.

The obsolescence of DVM creates a gap because all migration primitives (e.g., method or thread-based) based on DVM are incompatible with ART. Therefore, a platform-dependent process-level migration mechanism is required to enable application migration for future IoT applications that contain a bulk of native machine instructions. We utilize this concept for MCE\footnote{In this paper, we use the term mobile cloud environments (MCE) to generalize both MCC and MEC}. 

\subsection{Related Works and Contributions}
\label{rw}
Computational offloading has become one of the crucial problems in MCE due to the complexity and interdependencies of modern mobile/IoT applications, heterogeneity between mobile/IoT devices and the cloud/edge infrastructure, unpredictability, variability of wireless connectivity, and security issues. In this section, we discuss the computational offloading solutions proposed in the IoT-supported MEC/fog computing paradigm.

Bellavista  et al. \cite{8255754} proposed human-driven edge computing to ease the provisioning and extend the coverage of traditional MEC solutions for IoT and cyberphysical systems application scenarios. Similarly, a general framework for IoT fog-cloud applications, along with a delay-minimizing collaboration and offloading policy for fog-capable devices, was proposed in \cite{8246720}. This framework aimed to reduce the service delay of IoT applications. Lyu et. al. \cite{8270632} introduced an integration architecture of cloud, MEC, and IoT and proposed a request-and-admission framework to resolve the scalability problem. Without coordination among devices, the proposed framework separately operates at the IoT devices and computing servers by encapsulating the latency requirements in offloading requests. To meet the heterogeneity in the requirements of the offloaded IoT tasks in a multiaccess edge network, (e.g., different computing requirements and latency), Alameddine et al. \cite{8630994} mathematically formulated the dynamic task offloading and scheduling problem, which encompass three subproblems. The authors designed a decomposition method  based on logic-based bender decomposition to decide jointly on the task offloading (i.e., tasks to application assignment) and scheduling. 

Chen et al. \cite{8270633} devised a resource-efficient edge computing scheme in which an intelligent IoT device user can support the device's computation-intensive task through proper task offloading across local and nearby devices and the edge cloud in proximity. They explored the perspective of resource efficiency and proposed a delay-aware task graph partition algorithm and an optimal virtual machine selection method to minimize the edge resource occupancy of an intelligent IoT device by satisfying its quality of service (QoS) requirement. A density-based offloading strategy was analyzed in \cite{8542668}. The authors first utilized a strategy based on the density of IoT devices using k-means algorithm to partition the network of edge servers. Then, they analyzed and developed mathematical models to offload heterogeneous and uncertain tasks from various IoT devices to the edge server. An algorithm that utilizes the sample average approximation method for IoT devices was proposed, resulting in optimal computation offloading decisions.

The authors of \cite{8620684}  investigated the utilization problem of heterogeneous computation resources at the network edge to achieve the best energy efficiency among multiple-end devices while satisfying their delay requirements. They studied the computation offloading management problem by jointly considering the heterogeneous computation resources, latency requirements, power consumption at end devices, and channel states. The formulated energy minimization problem falls into the category of mixed-integer and nonlinear program. To solve the problem efficiently, the researchers decomposed the original problem into two subproblems and proposed an iterative solution framework that deals with transmission power allocation strategy and computation offloading. Elazahry et. al. \cite{8345635} formulated IoT offloading as a decision problem. They proposed a W5 reference framework for future research works. Cheng et. al \cite{8672604} investigated the computing offloading problem in a space-air-ground integrated network (SAGIN) and proposed a deep reinforcement learning-based computing offloading approach to learn the optimal offloading policy on the fly from dynamic SAGIN environments. They also proposed a joint resource allocation and task scheduling approach for unmanned aerial vehicle-based edge servers to allocate the computing resources to virtual machines and schedule the offloading tasks efficiently.

The authors of \cite{8360511} formulated a computational offloading game to model the competition between IoT users and efficiently allocate the limited processing power of fog nodes. They confirmed the existence of a pure Nash equilibrium and provided an upper bound for the price of anarchy. They further proposed a near-optimal polynomial time resource allocation mechanism wherein each user aims to maximize the net utility. The study conducted in \cite{8270634} introduced a generic fiber-wireless architecture that coexists with centralized cloud and distributes MEC for IoT connectivity. The problem of cloud-MEC collaborative computation offloading was defined, and a game-theoretic collaborative computation offloading scheme was proposed. Some early solutions proposed in MCC, such as Cloudlets \cite{5280678}, paranoid Android \cite{Paranoid}, virtual smartphone \cite{Virt:Phone}, and phone mirroring \cite{mirrorraey}, have utilized virtual machine migration primitives to harness the computational power on the edge. Similarly, the studies conducted in \cite{ThinkAir,simantaraey,raeyraey,6280375,Lee2012,Verbelen20111871,Verbelen20122629,raey,6834941,Wu2018} utilized code migration, which rely on programmers to specify program partitions using code annotation and skeletons. Meanwhile, CloneCloud \cite{Chun:CloneCloud} and its variants \cite{Yang2014}, as well as COMET \cite{gordon2012comet}, utilized the concept of thread migration.
Yousafzai et al. \cite{yousafzai2016computational} proposed a framework to offload ART-based mobile applications. The framework uses cloud data center support to enable offloading services for heterogeneous mobile devices. The researchers of \cite{neto2018uloof} presented a user-level online offloading framework, which is based on a decision engine that minimizes remote execution overhead without requiring any modification in the device's operating system. In summary, most computational offloading primitives strictly depend on virtualization technology.

The aforementioned research focuses on the formal modeling of computational offloading decision and optimization algorithms rather than focusing on process-level computational offloading. In addition, the existing offloading solutions do not consider the native code of IoT applications. However, modern mobile and IoT applications are compiled into native machine-dependent binaries upon their installation. Unlike the existing studies, the main contributions of this paper are summarized as follows:

\begin{enumerate}[label=\alph*)]
    \item We establish the case for process migration-based computational offloading in IoT-supported mobile edge/cloud computing.
    \item We propose a process migration-based computational offloading (PMCO) framework that transparently migrates a running process from a resource-constrained mobile device to resource-rich computing infrastructure. The proposed framework does not require an application source code on the edge server, which is a prominent case in MEC scenarios. 
    \item We evaluate the proposed framework and its lightweight features using standardized synthetic benchmarking experiments on an indigenous IoT research testbed measuring application execution time, amount of data transfer, and consumed energy.
\end{enumerate}

The rest of the paper is organized as follows:  Section II presents the components of the proposed PMCO framework. Section III discusses the details of the algorithmic procedures involved in the proposed framework. Section IV outlines the performance evaluation results, followed by the concluding remarks provided in Section V.

\section{Components of the proposed framework}
\label{sec_exp}

\begin{figure}
  \centering
      \includegraphics[scale=0.47]{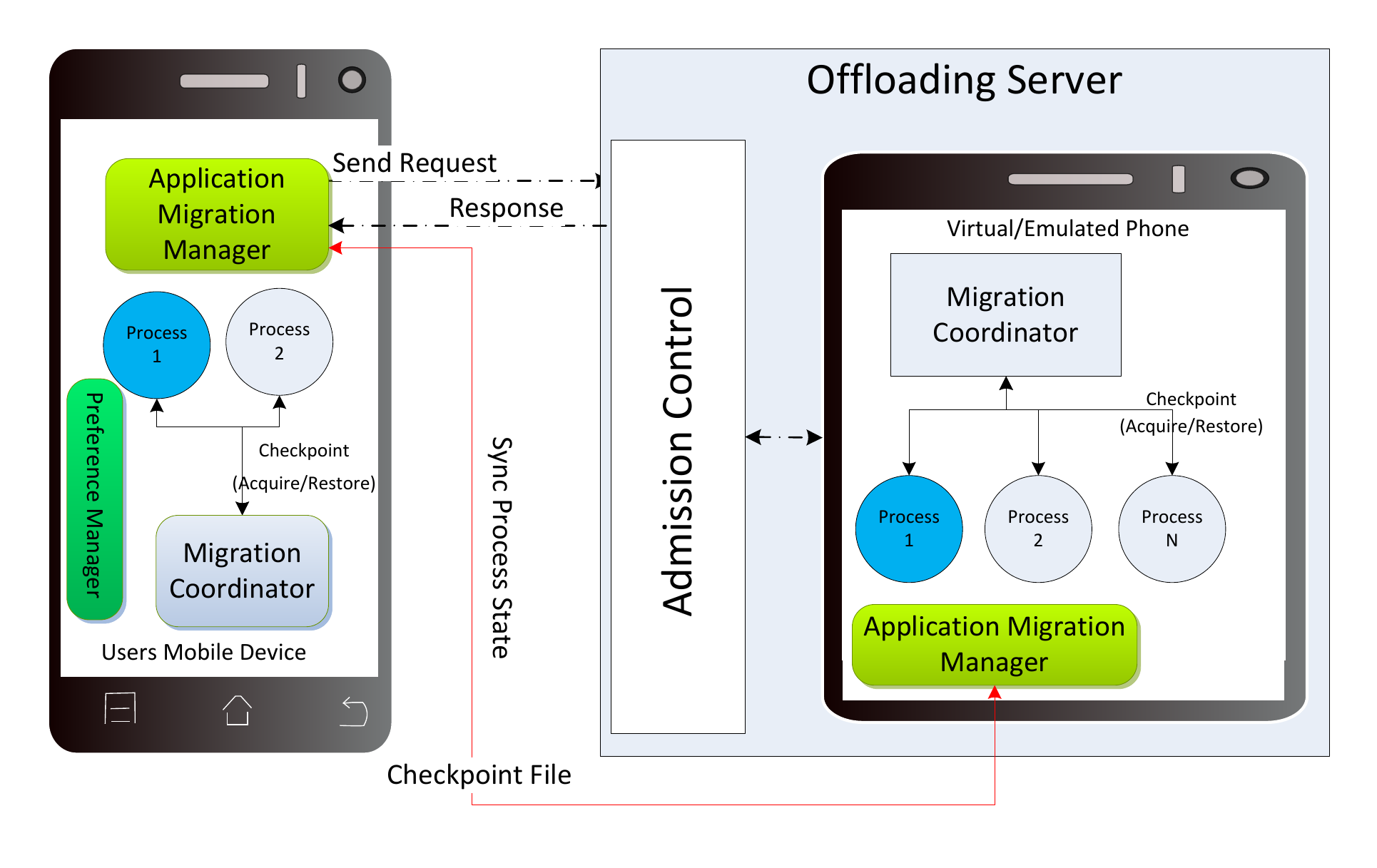}
  \caption{Bird eye view of proposed PMCO for mobile device augmentation.}
  \label{fig:framework}
\end{figure}

Our proposed framework consists of five major building blocks, namely, user-side migration preference manager (UMPM), user-side application migration manager (UAMM), user-side application migration coordinator (UAMC), edge-side admission control (EAC), and edge-side application migration manager (EAMM), as shown in figure \ref{fig:framework}. The details of these components are provided in the following subsections.

\subsection{UMPM}
UMPM allows users to enable/disable runtime application offloading preferences for mobile/IoT applications installed in the client's device. This preference manager helps avoid the computationally expensive static analysis of user applications. Preferences are stored in a key/value-based fashion in the preference registry. The preferences defined in this registry are classified into two categories: global and application-specific preferences.
\subsubsection{Global preferences} 
Global preferences apply to all mobile applications. These preferences include parameters, such as $E_c, E_i,E_t,E_r,S_{m},B_t$,  and server selection criteria (cost per service), where $E_c$ stands for active energy consumption of the device's CPU; $E_i$ is the idle energy consumption of the device's CPU; $E_t$ and $E_r$ correspond to energy consumption while the device's wireless interface is in transmission and reception mode, respectively; $S_{m}$ and $S_{c}$ represent the computing power (in MIPS) of the mobile device and remote edge server, respectively; $\beta_{u}$ and $\beta_{d}$ represent the network's uplink and downlink bandwidth; and $B_t$ is the benefit threshold. Bt indicates the energy-saving amount of the computational offloading process required by the offloading decision function (Equation \eqref{eq:ch4:energy_saving})  to trigger the offloading process.

\begin{equation}
\label{eq:ch4:energy_saving}
	[[E_{c} \times \frac{I}{S_{m}}] - [E_i \times \frac{I}{S_{c}}] - [E_t \times \frac{\alpha}{\beta_u}] - [E_r \times \frac{\gamma}{\beta_d}]] > B_t , where B_t \geqslant 0
\end{equation}

\subsubsection{Application specific preferences}
Application-specific parameters are defined as a $4$-tuple ($<I,\alpha,\gamma,P_f>$) for all applications that can be offloaded. $I$ is the number of instructions in the application; $\alpha$ represents the amount of data that will at least be transferred in a single offloading transaction ($\alpha$ is the only value that the offloading engine can accurately obtain without any overhead because it can be calculated at the time of offloading); and $\gamma$  is the amount of data that will be received in a single offloading transaction. Intuitively, $\gamma$  is unknown at the time of offloading decision without prediction, estimation, or pre-execution, thus becoming expensive and counterproductive because of the increase in the offloading decision time. Therefore, the preference manager simplifies the offloading process with user assistance. $P_f$  is a binary variable that can be set to enable forced migration or disable migration for a specific application.

\subsection{UAMM}
UAMM is responsible for discovering, selecting, and interfacing the edge service provider. As discussed in the previous subsection, an edge service is selected on the basis of the preferences (cost and QoS metrics) defined by the mobile user and the offloading decision function. After selecting the service and establishing a connection, the UAMM initiates the UAMC. Then, the UAMM iterates over the UMPM application list. If an application needs to be offloaded and obtains a positive result from (Equation \eqref{eq:ch4:energy_saving}), UAMM will launch the application using UAMC. The UAMM will wait for a signal from the UAMC about a checkpoint package, which needs to be shipped to a remote device on the edge infrastructure. The UAMM will then fetch the checkpoint package from the local temporary storage designated for the process and dispatch the checkpoint along with a meta information. The meta information specifies the size, as well as whether the application is migration aware or does not contain any migration marker. This meta information is utilized by the EAMM to restart the checkpoint image correctly. Once the checkpoint package is transferred to the edge device, UAMM will wait for the response from the edge device. Once the response package, which is also a checkpoint image, is received from the edge, UAMM will restart the received checkpoint file on the device.

\subsection{UAMC}
UAMC is a modified over-the-counter checkpoint-restart  tool\footnote{Accessed on: 25 February 2019  \url{https://github.com/dmtcp/dmtcp}}, which can checkpoint a running process and subsequently restart it on a remote computing infrastructure or mobile device. Executing an application through UAMC launches a coordinator process on the local client's device to assist the migration process \cite{ansel2009dmtcp}. The coordinator is stateless; thus, when it crashes and subsequently reboots, the process that the coordinator needs to checkpoint is also restarted. The coordinator uses two approaches to send the checkpoint signals to the running applications. The first approach is applied to applications without migration markers (i.e., nonmigration-aware applications). In MCE, such applications are referred to as unmodified applications that do not have programmer-defined migration markers. Unmodified applications cannot checkpoint themselves. The second approach applies to applications that can be modified (either by the developer or automatically during the compilation process) to include migration markers. Applications with migration markers can readily checkpoint themselves. The former is resource-hungry because it may checkpoint the application multiple times, whereas the latter is lightweight due to its ability to checkpoint upon reaching a marker.

The checkpoint mechanism interrogates kernel state (open file descriptors, file descriptor offset) and saves register values using available kernel data structures. Lastly, all user-space memory is stored to the checkpoint image (/proc/self/maps), which includes any library that the process is using. This strategy improves the portability of the checkpoint images and even allows the migration of the process in some cases to hosts with different Linux distributions and kernels. The checkpoint mechanism is configured to use gzip for the compression of checkpoint images, resulting in the reduction of network traffic.

\subsection{EAC}
 \label{sec:adm:control}
 The admission control unit imposes the fair-share usage policy for edge resources. In the case of an edge device that has a weak pricing model admission control, the admission unit enforces the efficacy of the system \cite{7295892}. Furthermore, offloading in MEC faces scalability problems due to numerous mobile and IoT devices \cite{8270632}. Admission control manages user request to resolve the scalability, security, and load balancing problem in the edge infrastructure \cite{8360851}. The administration depends on the edge provider policy. User requests are queued and processed using any state-of-the-art scheduling discipline defined by the edge provider \cite{8253470}. Whenever an IoT device sends a task offloading request to the edge server, the admission control verifies whether the device can be emulated/virtualized or not. In addition, the admission control also ensures the availability of any software platform that is required for the offloaded task. If both conditions are satisfied, the request is granted, and the subsequent process state synchronization packets from that device are directly transmitted/forwarded to the emulated/virtualized edge instance that manages users' request.

\subsection{EAMM}
EAMM resides in a virtualized/emulated or physical device instance in the edge infrastructure. EAMM acts as a server to the UAMM residing on the user's mobile/IoT device. To maintain the policies implemented by the admission control, EAMM tracks the service time of the user requests. If the admission control policy is violated, EAMM stops servicing the mobile device. Once EAMM receives an offloading package from the mobile device, it restarts the application package. Then, once the application is restarted on the edge device, EAMM recheckpoints the application within regular intervals for applications without migration markers. Meanwhile, if the application is migration aware, the application recheckpoints itself after reaching a certain marker. The recheckpoint of the application at the edge is performed to transfer the computation and execution control back to the mobile device.

\section{Proposed PMCO Algorithms}
In this section, we present our proposed algorithms, which originally correspond to the interaction between the components of the drafted framework detailed in Section \ref{sec_exp}. The implementation of PMCO can vary depending on checkpoint methodology, hardware, and software architecture. The primary steps in the generic process migration can be found in \cite{vasudevandesign}. Based on these generic steps and the components of our proposed solution, we present the PMCO algorithm and an edge-side service algorithm to serve the IoT end users.

\begin{figure}[h!]
\caption*{\textbf{Algorithm 1}: Offloading Algorithm}
\label{algo:offloadEngine}
\begin{algorithmic}[1]
\STATE $E_c, E_i,E_t,E_r,S_{m},B_t \gets$ MPMGetGlobalPrefrences()
\STATE $S \gets$ AMMGetFindServer()
\STATE $\beta_u,E_t \gets$ TestUpload()
\STATE $\beta_d,E_r \gets$ TestDownload()
\WHILE{ isConnected(S) } 
\STATE $P,I,\alpha,\gamma,P_f,P_t \gets$ MPMGetProcessInfo()
\IF{ $P \neq NULL$} 
\STATE $AMC \gets$ LaunchApplicationMigrationCoordinator \phantom a \phantom . \phantom . \phantom . \phantom . \phantom .  ($E_c,E_i,E_t,E_r,\beta_u,\beta_d,\alpha,\gamma$)
\STATE LaunchProcess($P$)
\IF{ $P_t = $ MigrationAware} 
\STATE $O_t \gets$ AMMInfoPolling()
\IF{ $O_t  \neq $ NULL} 
\STATE $\beta_u \gets $ TransferCheckpoint($O_t,P_t,S$)
\STATE $\beta_d \gets $ ReceiveCheckpoint($O_r,S$)
\STATE RestartCheckpoint($O_r$)
\ENDIF
\ELSE
\IF{ Benefit($E_c,E_i,E_t,E_r,\beta_u,\beta_d,\alpha,\gamma$) $ >  0$) $\OR P_f = 1 $} 
\STATE AMMSignalCheckpoint($P$)
\STATE $O_t \gets$ AMMInfoPolling()
\STATE AMMSignalKill($P$)
\IF{ $O_t  \neq $ NULL} 
\STATE $\beta_u \gets $ TransferCheckpoint($O_t,P_t,S$)
\STATE $\beta_d \gets $ ReceiveCheckpoint($O_r,S$)
\STATE RestartCheckpoint($O_r$)
\ENDIF
\ELSE 
\STATE ExecuteProcessLocally($P$)
\ENDIF
\ENDIF
\ENDIF
\ENDWHILE
\WHILE{isNotConnected(S)} 
\STATE $P \gets$ MPMGetProcessInfo()
\IF{ $P \neq NULL$} 
\STATE ExecuteProcessLocally($P$)
\ENDIF
\ENDWHILE
\end{algorithmic}
\end{figure}

Algorithm 1  begins with initializing the global parameter values defined by the IoT user in the UMPM registry (Line 1). The UAMM searches and negotiates for a suitable edge server according to the criteria defined by the IoT user (Line 2). Once the mobile device is connected to the edge server, a test upload and download transmission is performed (Lines 3-4). Test transmission is performed to gather the upload bandwidth $\beta_u$ and the download bandwidth $\beta_d$ along with $E_t$ and $E_r$, if not set by the IoT user in the global preferences. Subsequently, the main offloading loop, which is conditioned by the server connectivity, is executed (Lines 5-32). The UMPM registry is queried to obtain an offloading candidate application that is listed along with its preferences (Line 6). Once an application has been fetched, the application is executed through UAMC, along with the parameters required for making an offloading decision (Lines 8-9). Afterward, the type of application is determined (i.e., whether the application is migration aware or not) (Line 10). If the application is migration aware, the offloading decision benefit function (equation \eqref{eq:ch4:offloadingDecesion}) is called inside the application execution life cycle at the point of offloading. If the function returns true, the application is checkpointed, and UAMC is acknowledged about the checkpoint. The acknowledgment means that a checkpoint is ready and must be offloaded to the remote device (Line 13). In retrospect, the algorithm waits to receive the updated checkpoint from the remote device (Line 14). Meanwhile, if the application is not migration aware, the control is then jumped to where the offloading algorithm calls the benefit function listed in equation \eqref{eq:ch4:offloadingDecesion} (Line 18). If the offloading benefit function returns true, then the UAMM sends signals for the UAMC to send a checkpoint signal to the process checkpoint thread (Line 19). Once the process receives that signal, the process checkpoints itself and acknowledges the UAMC, which is handled by the polling function (Line 20). Once the acknowledgment is received, the process is terminated, and the checkpoint is transferred to the remote device (Line 23). In response, the algorithm waits for receiving the updated checkpoint from the remote device (Line 24). If the benefit function (Line 18) returns false, then the application is locally executed (Lines 33-37). Lastly, if the mobile device is not connected to an edge service, the application is locally executed on the mobile/IoT device, irrespective if a UMPM entry for the mobile application is available and true.
\\

Equation \eqref{eq:ch4:offloadingDecesion} presents an offloading decision function, which is modified from Equation (1) to consider the PMCO. This equation needs to be modified for additional decision criterion.

\begin{equation}
\label{eq:ch4:offloadingDecesion}
	[E_{m}^p  - E_{m}^{'p}  - E_{m}^{''p} - E_{t}^p - E_{r}^p ]  > B_t, where B_t \geqslant 0
\end{equation}

where $E_{m}^p$ represents the energy consumption of process $p$  on the source mobile device $m$, $E_{m}^{'p}$ is the energy consumption of the checkpointing of process $p$ on the source mobile device $m$, $E_{m}^{''p}$ represents the energy consumption of restarting the process from a checkpoint state (received from a remote computing device) on the on-the-source mobile device $m$. $E_{t}^p$ and $E_{r}^p$ represent the energy consumption required to transfer and receive a checkpoint of process p, respectively. The modification of the decision function is necessary for process migration in which the process is terminated after the checkpoint and subsequently transferred to the remote device where it is restarted on the source device by considering the updated state.\\

In Algorithm 2, we present the step sequence of the edge-side offloading service provided to the IoT end users. 

\begin{figure}[h!]
\caption*{\textbf{Algorithm 2}: Offloading Service}
\label{algo:offloadService}
\begin{algorithmic}[1]
\WHILE{1} 
\STATE $C \gets$ AcceptConnection()
\STATE ReceiveCheckpoint($O_r,P_t,C$)
\IF{ $P_t = $ MigrationAware} 
\STATE $P \gets $ RestartCheckpoint($O_r$)
\STATE $O_t \gets$ AMMInfoPolling()
\IF{ $O_t  \neq $ NULL} 
\STATE TransferCheckpoint($O_t,S$)
\ENDIF
\ELSE
\STATE $P \gets $ RestartCheckpoint($O_r,Interval$)
\IF{ isProcessExecutionFinished(P)} 
\STATE $O_t \gets$ AMMInfoPolling()
\IF{ $O_t  \neq $ NULL} 
\STATE TransferCheckpoint($O_t,S$)
\ENDIF
\ENDIF
\ENDIF
\ENDWHILE
\end{algorithmic}
\end{figure}

Client connection is established, and then a checkpoint file is accepted (Lines 2-3). Once the checkpoint file is received, the service checks if the application for which the checkpoint file is received is migration aware (Line 4). If the received application is migration aware, then the application is restarted with its own checkpoint coordinator process (Line 5). Then, the EAMM starts polling to prepare a new checkpoint image that will be sent to the client device (Line 6). If the received application is not migration aware, the received application is restarted with its own coordinator on an interval-based checkpoint (Line 11). Once the received application is restarted, the offloading service will wait until the process is finished. The offloading service conducts a poll (Line 13) of the latest checkpoint image, which is generated on the basis of the interval timeout during program execution. Finally, the checkpoint image is returned to the client's device for re-execution.

\section{Performance evaluation results}
\label{evaluation}
This section presents the performance evaluation of our proposed framework. To evaluate the proposed model and its lightweight features, we use standardized synthetic benchmarking experiments with a computation-intensive application for measuring mega floating point operation per second, mega instruction per second, application execution time, amount of data transfer, and consumed energy. A total of 30 observations for each benchmark application was collected and analyzed.

\subsection{Experimental setup}
\label{chp5:ec}

To benchmark the proposed prototype, we selected four standard synthetic computing benchmarks and a synthetic computation-intensive application with various execution input granularities. The selected synthetic benchmarks included (i) Dhrystone\footnote{Accessed on: 28 February 2019 \url{https://en.wikipedia.org/wiki/Dhrystone}}, (ii) Whetstone\footnote{Accessed on: 28 February 2019 \url{https://en.wikipedia.org/wiki/Whetstone_(benchmark)}}, (iii) Linpack\footnote{Accessed on: 28 February 2019 \url{https://en.wikipedia.org/wiki/LINPACK}}, and (iv) Scimark2\footnote{Accessed on: 28 February 2019 \url{http://math.nist.gov/scimark2/}}. The Scimark benchmark was configured to execute the large instance; the configuration can be set as a command line parameter using the source code provided by National Institute of Standards and Technology. Table \ref{tab:ch5:gran} presents the computation-intensive application, which is a matrix multiplication application with different matrix granularities. All applications (benchmarks and matrix multiplication programs) were annotated with migration points that enable self-checkpointing. These applications were compiled using a standard GNU compiler collection (GCC) with position independent codes (-fPIC) switch. -fPIC is required by the checkpointing engine to checkpoint and transparently restart the process in user space.

\begin{table}[]
\centering
\caption{Benchmark Matrix Multiplication Granularity }
\label{tab:ch5:gran}
\begin{tabular}{@{}l|l@{}}
Workload\#1 & Matrix Granularity \\ \toprule
1           & 300x300              \\  \midrule
2           & 400x400            \\ \midrule
3           & 500x500            \\ \midrule
4           & 600x600            \\ \midrule
5           & 700x700            \\ \midrule
6           & 800x800            \\ \midrule
7           & 900x900            \\ \midrule
8           & 1000x1000          \\ \bottomrule
\end{tabular}
\end{table}

\begin{figure}[!htb]
  \centering
      \includegraphics[scale=0.7]{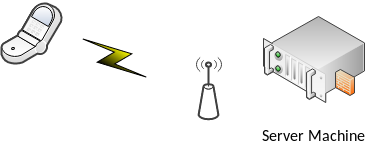}
\caption{Network topology of experimental setup.}
  \label{fig:ch5:EC}
\end{figure}

The primary data for the performance evaluation were collected by testing the prototype applications in three different real scenarios. In the first scenario, all components of the mobile/IoT applications were executed on a local mobile device to analyze the performance evaluation metrics of the application. In the second scenario, the application was again executed on local mobile/IoT device using the proposed framework to analyze the framework overhead. In the third scenario, the components of the mobile/IoT application were offloaded at runtime by implementing the proposed computational offloading techniques. A schematic of our benchmarking setup is shown below (figure \ref{fig:ch5:EC}).

\begin{figure*}[!htb]
   \centering
       \subfloat[][]{\includegraphics[width=.4\textwidth]{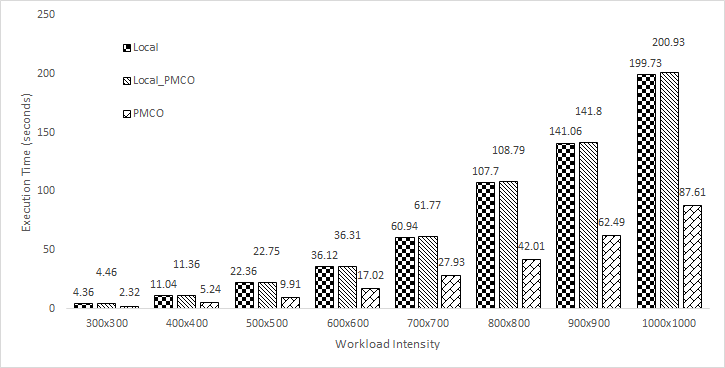}}\quad
   \subfloat[][]{\includegraphics[width=.4\textwidth]{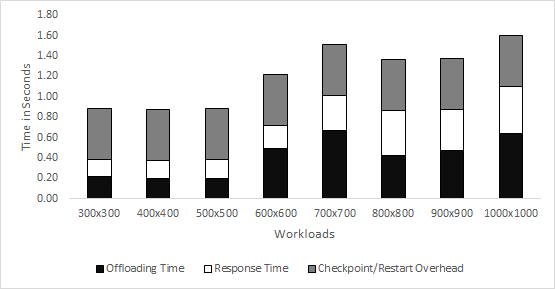}}\\
   \subfloat[][]{\includegraphics[width=.4\textwidth]{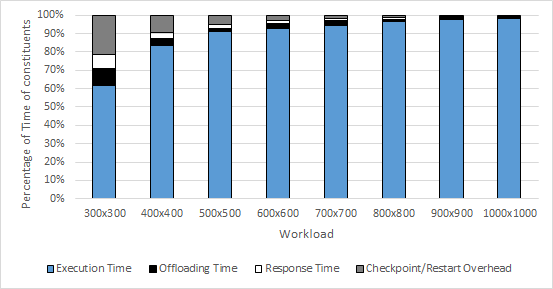}}\quad
   \subfloat[][]{\includegraphics[width=.4\textwidth]{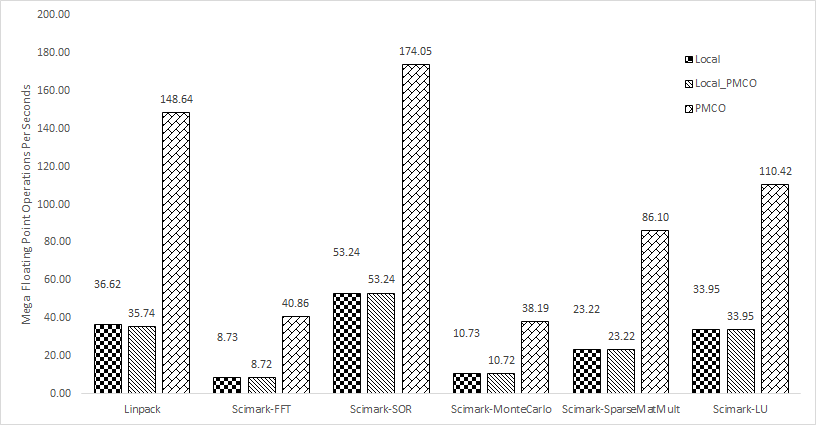}}
   \caption{(a) Execution time for matrix multiplication workloads generated via experimentation. (b) Breakdown of the contributing factors of remote execution time using PMCO. (c) Impact of the contributing factors on remote execution time using PMCO. (d) Mean MFLOPS values for 30 observations of Linpack and Scimark benchmarks generated via Local, Local\char`_PMCO and PMCO executions.}
   \label{fig:thirdMulti1}
\end{figure*}

The process state is platform dependent. To prepare a real experimentation environment, we needed similar hardware and operating system architecture on both endpoints of the experimental setup, as shown in figure \ref{fig:ch5:EC}. We used a Samsung Galaxy S-II i9100g, which is considered a core component of the IoT. This device is equipped with a dual core 1.2 GHz TI OMAP 4430 ARM Cortex-A9 SoC and 1 GB RAM. The client device is connected to the edge server device using the campus WiFi network. The edge server device is a high-performance Sony Xperia Z Ultra equipped with a quad-core 2.2 GHz Qualcomm MSM8274 ARM Cortex-A9 SoC. To differentiate the client and the server device further, we down-clocked the client device from 1.2 GHz to 600 MHz to emulate a resource-poor client device and a resource-rich server device.
Both devices were based on Android OS. In our experimental setup, we utilized the Android kernel but deployed a standard ARM port of Ubuntu 13.10 Saucy Salamander in a chroot environment over both devices. This chroot jail over Android allowed us to utilize the full capabilities of the standard GNU Glibc (missing in Android) and Android kernel (Linux kernel), especially in accessing the process state from user spaces.

The application migration manager was programmed in Java to provide a client-server communication interface between both endpoints, whereas for checkpoint management and coordination, we used and modified distributed multithreaded checkpointing (DMTCP)\footnote{Accessed on: 25 February 2019 \url{http://dmtcp.sourceforge.net/}}, a multipurpose checkpointing mechanism for parallel and distributed computing environments. DMTCP is community driven and has a progressive enhancement; the framework can be ported to new platforms and environments, thereby enabling portability to a large extent.

We evaluated the performance of each benchmark and the eight different granularities of the matrix multiplication workload in three modes, namely, Local, Local\char`_PMCO, and the proposed PMCO. Each benchmark and matrix multiplication workload were executed 30 times, and the mean value was considered for analysis. To enhance the reliability of the results and ensure that the data collection is unbiased, we presented the findings with 95\% confidence interval.

The improvement in computing power/execution time was measured by executing the benchmark applications, which are specifically designed for benchmarking computing systems and hardware. The migration segments of the benchmark applications comprise the start and end of the main function of the application to ensure that the generated benchmark values are unbiased and can provide the real increase in computing power based on the edge device. The benchmark execution results were automatically encapsulated in the checkpoint parcel by the checkpoint manager (EAMM) on the edge server.

To collect the energy consumption of the complete application under execution, we monitored the application and the framework components using PowerAPI \cite{bourdon2013powerapi}in console mode, which is a granular tool for investigating the power consumption of running applications based on the chipset thermal design power (TDP). The TDP value for the client device (Samsung Galaxy SII i9100g) was set to 0.6 W\footnote{Accessed on: 30 January 2019 \url{http://www.notebookcheck.net/Texas-Instruments-TI-OMAP-4430-SoC.86865.0.html}}. PowerAPI also provides a feature to monitor the aggregate energy consumption of a group of processes, which in our scenario of remote execution (i.e., process migration) was configured to monitor the power consumption of the UAMM and the UAMC components of the proposed framework. Energy consumed by other software components were not considered in this data collection phase, and the applications in active mode were run throughout the experiment by preventing any pause of block in the application execution. Furthermore, we parsed the power profile output generated by PowerAPI using bash scripts to generate the CSV files for analysis.

The improvement in execution time was measured using the matrix multiplication applications because benchmarks were not designed for this purpose (i.e., benchmarks will run for the same time units on all devices to provide an unbiased analysis of the computing capabilities). The execution time of the matrix multiplication application was measured by the application itself because we enclosed the migration segment of the source code with timers to provide details about the execution time during local or remote execution. Thus, the execution time of the program on the platform where it is executing was provided. The remote execution time also contained the checkpoint restart overhead and the transmission and reception time, and their respective components generated these values in a formatted log file, which was then processed using bash scripts to generate the CSV files for analysis.

Our last parameter is the amount of data transfer or the size of checkpoint file, which was first migrated from the mobile device. An updated checkpoint state was transferred by the mobile device from the server. The amount of data transfer was logged by the UAMM to provide the bandwidth consumption details. The log file was processed similarly as the other parameters.

%\begin{figure*}[!h]
%  \centering
%      \includegraphics[scale=0.75]{BarChartMatrixWorkload.png}
%  \caption{Execution time for matrix multiplication workloads generated via experimentation.}
%  \label{chap6:fig:execution_time}
%\end{figure*}

\subsection{Execution time and Computing Power}

\begin{figure}[h]
  \centering
      \includegraphics[scale=0.6]{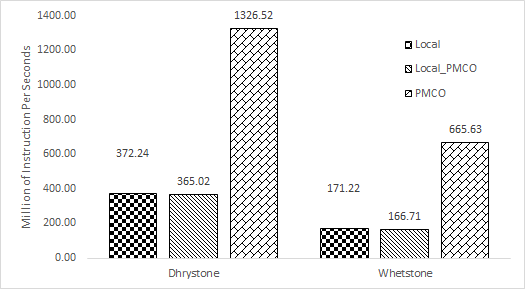}
  \caption{Mean MIPS and MWIPS values for 30 observations of Dhrystone and Whetstone benchmarks generated via Local, Local\char`_PMCO and PMCO executions.}
  \label{chap6:fig:compute_power:dwstone}
\end{figure}

\begin{figure}[h]
  \centering
      \includegraphics[scale=0.5]{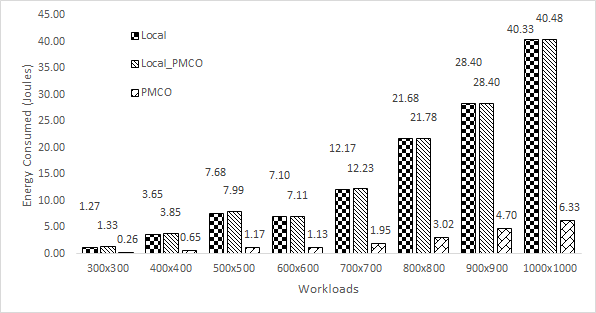}
  \caption{Energy consumed for matrix multiplication workloads gathered via PowerAPI.}
  \label{chap6:fig:consumed_energy}
\end{figure}

This section presents the temporal results of the execution of the experimental application of matrix multiplication and the benchmarking applications in three environments. One environment was local, in which the entire application-including intensive and nonintensive-was executed on the mobile device. In the second environment (Local\char`_PMCO), the experimental application of the matrix multiplication was locally executed through the PMCO environment. The third environment was the proposed PMCO in which application execution started locally. When the execution reached a migration marker, the process was offloaded to the mobile cloud for execution. Data related to execution were gathered using experimental analysis. Several charts were used to demonstrate the findings.

\begin{figure*}[h]
   \centering
       \subfloat[][]{\includegraphics[width=.4\textwidth]{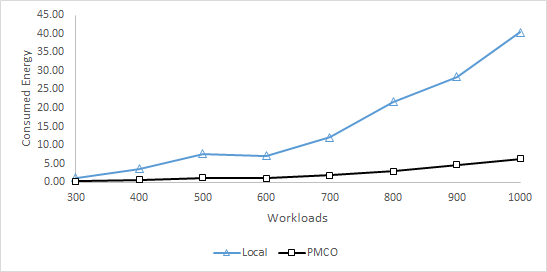}}\quad
   \subfloat[][]{\includegraphics[width=.4\textwidth]{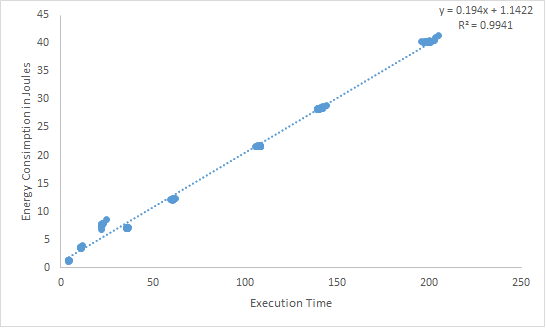}}\\
   \subfloat[][]{\includegraphics[width=.4\textwidth]{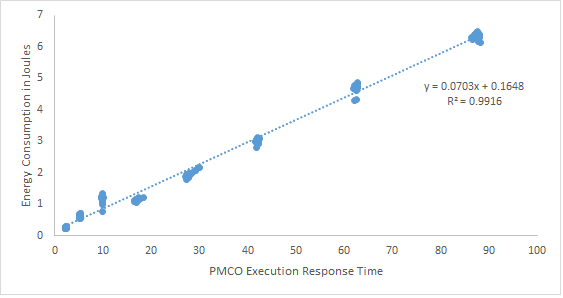}}\quad
   \subfloat[][]{\includegraphics[width=.4\textwidth]{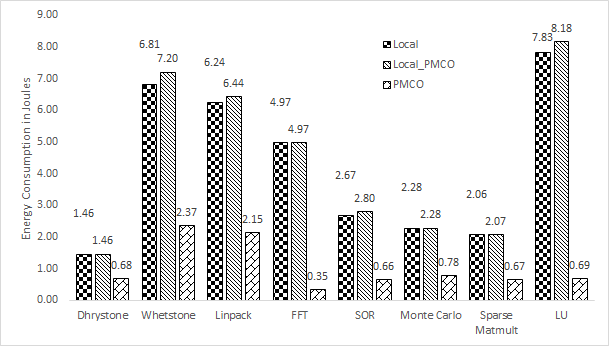}}
   \caption{(a) Scattered plot with interpolation lines for matrix multiplication energy consumption. (b) Scatter plot showing linearity correlation between local execution time and consumed energy. (c) Scatter plot showing linearity correlation between PMCO execution response time and consumed energy. (d) Energy consumed for benchmark application workloads gathered via PowerAPI.}
   \label{fig:secondMulti}
\end{figure*}

The mean execution time of the workloads in PMCO execution mode was 1.88, 2.10, 2.25, 2.12, 2.18, 2.56, 2.25, and 2.27 times  better than the local execution. Each diagonal brick bar in figure \ref{fig:thirdMulti1}(a) represents the mean value of the execution time measured by PMCO for 30 iterations of each matrix multiplication workload. Similarly, each diagonal strip bar represents the mean execution time measured using Local\char`_PMCO, whereas each checkered bar represents the execution time for local execution. Figure \ref{fig:thirdMulti1}(b) depicts the increase in complexity as the matrix multiplication intensity increases from left to right. The growth of the workload intensity had a significant effect on execution time when the workloads were entirely executed on the mobile device. Meanwhile, the execution time in PMCO execution mode was remarkably smaller than local execution. The execution of the last workload on the local client's device took more than 199 s to complete, which suggests the incapacity to execute high intensity workloads on the client device.

The improvement in execution time is not a function of intensity level of the workload but is a function of the computing power of the remote server and the underlying network conditions. The mean execution time savings according to the increasing order of workload intensity are 53\%, 47\%, 44\%, 47\%, 45\%, 39\%, 44\%, and 43\%. Data transmission time plays a vital role in the overall execution response time of the workload. Similarly, from the overhead point of view, when the workloads were locally executed through the PMCO framework, the degradation in increasing order of workload intensities was 2.16\%, 2.81\%, 1.73\%, 0.50\%, 1.32\%, 0.99\%, 0.51\%, and 0.59\%. The minimum degradation was approximately 0.50\%, and the maximum reached up to 2.81\%. However, we could not identify a correlation between the workload intensity level and the percentage of degradation because of the influences of the internal operating system threading model and the thread context switching. The correlation between the workload intensity and time saving is illustrated in figure \ref{fig:thirdMulti1}(a), indicating that low workload intensities have smaller discrepancies compared with higher ones.

The execution time in local mode is highly affected by the workload intensity and the computing power of the mobile device (e.g., CPU clock speed, RAM, storage, and cache). However, the PMCO execution time also depends on other metrics, such as checkpoint/restart time and underlying network bandwidth. These contributing factors play a crucial role in the overall offloading benefit. Figure \ref{fig:thirdMulti1}(b) shows a stacked chart with the breakdown of the details in PMCO execution mode. An increase in the offloading time (from the $500 \times 500$ to the $600 \times 600$) was observed. Similarly, a decrease in offloading time (from $700 \times 700$ to the $800 \times 800$) was discerned. Furthermore, the number of the $800 \times 800$ and the $900 \times 900$ matrix workload offloading times were less than the $600 \times 600$ ones. The variation of results is due to the fact that at a certain point, the campus network bandwidth fluctuates, thereby affecting the data rate at which the checkpoint image is transferred and eventually increasing the offloading time and the total execution time. Figure \ref{fig:thirdMulti1}(c) shows a 100\% stacked chart, which shows that in the first workload, the overhead of the contributing factors was approximately 40\%. As the workload intensity increases, the overhead of the contributing factors decreases. The solid black color in the bar represents the offloading time. The white, gray, and blue chunks represent the response time, the checkpoint restart overhead, and the process execution time on the remote server, respectively.

%\begin{figure*}[!h]
%  \centering
%      \includegraphics[scale=0.75]{BarChartMatrixWorkload.png}
%  \caption{Execution time for matrix multiplication workloads generated via experimentation.}
%  \label{chap6:fig:execution_time}
%\end{figure*}

%\begin{figure*}[!htb]
%  \centering
%      \includegraphics[scale=0.9]{OverheadTimes.png}
%  \caption{Breakdown of the contributing factors of remote execution time using PMCO.}
%  \label{chap6:fig:execution_time_breakdown1}
%\end{figure*}

%\begin{figure*}[!htb]
%  \centering
%      \includegraphics[scale=0.9]{TimePercentagesStacked.png}
%  \caption{Impact of the contributing factors on remote execution time using PMCO.}
%  \label{chap6:fig:execution_time_breakdown2}
%\end{figure*}

%\begin{figure*}[!htb]
%  \centering
%      \includegraphics[scale=0.7]{BarChartScimarkLinpack.png}
%  \caption{Mean MFLOPS values for 30 observations of Linpack and Scimark benchmarks generated via Local, Local\char`_PMCO and PMCO executions.}
%  \label{chap6:fig:compute_power:linpack_scimark}
%\end{figure*}

%\begin{figure*}[!htb]
%   \centering
%       \subfloat[][]{\includegraphics[width=.4\textwidth]{BarChartDWStone.png}}\quad
%   \subfloat[][]{\includegraphics[width=.4\textwidth]{BarChartMatrixWorkloadEnergyConsumed.png}}
%   \caption{(a) Mean MIPS and MWIPS values for 30 observations of Dhrystone and Whetstone benchmarks generated via Local, Local\char`_PMCO and PMCO executions. (b) Energy consumed for matrix multiplication workloads gathered via PowerAPI.}
%   \label{fig:fourthMulti1}
%\end{figure*}

The data plotted in figures \ref{fig:thirdMulti1}(d) and \ref{chap6:fig:compute_power:dwstone} are the average computing power measured from the benchmark workloads for local, Local\char`_PMCO, and PMCO execution modes. Each diagonal brick bar represents the mean value of the computing power measured using PMCO mode in 30 iterations for each corresponding benchmark workload. Diagonal strip bars represent the mean computing power measured using the Local\char`_PMCO mode of execution, whereas the checkered bar represents the corresponding computing power for local execution. In summary, figures \ref{fig:thirdMulti1}(d) and \ref{chap6:fig:compute_power:dwstone} depict the increase in computing power that is disposed to the local mobile device through PMCO. The figure also demonstrates that a minimal margin of degradation is present if a benchmark is locally executed using PMCO components.

%\begin{figure}[!htb]
%  \centering
%      \includegraphics[scale=0.6]{ScatterLocalPMCOEnergyConsumed.png}
%  \caption{Scattered plot with interpolation lines for matrix multiplication energy consumption.}
%  \label{chap6:fig:consumed_energy_scatter}
%\end{figure}

%\begin{figure}[!htb]
%  \centering
%      \includegraphics[scale=0.6]{ScatterPlotExecutionTimeEnergyConsumed.png}
%  \caption{Scatter plot showing linearity correlation between local execution time and consumed energy.}
%  \label{chap6:fig:reg:energy_consumed_local:linear}
%\end{figure}

%\begin{figure}[!htb]
%  \centering
%      \includegraphics[scale=0.6]{ScatterPlotPMCOTimeEnergyConsumed.png}
%  \caption{Scatter plot showing linearity correlation between PMCO execution response time and consumed energy.}
%  \label{chap6:fig:reg:energy_consumed_pmco:linear}
%\end{figure}

%\begin{figure*}[!htb]
%  \centering
%      \includegraphics[scale=0.7]{BarChartBenchmarkEnergyConsumption.png}
%  \caption{Energy consumed for benchmark application workloads gathered via PowerAPI.}
%  \label{chap6:fig:consumed_energy_b}
%\end{figure*}

\subsection{Energy Consumption}
In this section, we present the energy consumption required in the execution of the experimental matrix multiplication application in the Local, Local\char`_PMCO, and PMCO execution modes along with their statistical comparison.

Figure \ref{chap6:fig:consumed_energy} depicts increasing energy consumption as the matrix multiplication intensity increases from left to right. The growth of the workloads had a significant effect on consumed energy when the workloads were entirely executed on the mobile device. The energy consumption using PMCO was remarkably smaller than that of the local execution. The execution of the last workload in the experiment took more than 40.33 J to complete, which suggests incapacity to execute intensive workloads on client devices.

Moreover, executing workloads on the local device in increasing order of workload intensity consumes 4.94, 5.62, 6.55, 6.27, 6.25, 7.16, 6.04, and 6.37 times  more energy compared with the PMCO execution. The mean energy consumption for the eight workloads was approximately 6.15 times more than the energy consumption of the application when it was executed on the client's device. For instance, if the energy required to run the eighth workload locally is 40.33 J, the same workload will consume as low as 6.33 J  when performed outside the mobile device. In terms of the overhead, when the workloads were locally executed through the PMCO framework, the degradation was approximately 0.01\% to 5\%. The data plotted in figure \ref{chap6:fig:consumed_energy} are the mean consumed energy of the workloads for Local, Local\char`_PMCO, and PMCO execution modes. Each diagonal brick bar represents the mean value of the consumed energy measured using PMCO, whereas each diagonal strip and checkered bar represent the mean consumed energy measured using the Local\char`_PMCO mode and local execution, respectively.

Significant differences in local and PMCO energy consumption enable mobile users to offload using the PMCO of extremely intensive workloads on their mobile devices to gain battery lifetime. These differences are depicted in figure \ref{fig:secondMulti}(a). Scattered triangles and squares across the graph and the corresponding interpolating lines show the differences in achievements and the correlation between the workload intensity and time saving, respectively. Workloads with low intensity have smaller discrepancies compared with those with high intensity (represented by the circle and the triangle). The increase in energy consumption of PMCO was not abrupt compared with local execution.

The energy consumption in local execution mode was highly affected by workload intensity, along with the power rating of the mobile device. This phenomenon is presented by the scattered plots in figures \ref{fig:secondMulti}(b) and \ref{fig:secondMulti}(c), which demonstrate that if workload intensity increases, the energy consumption also increases. Figure \ref{fig:secondMulti}(d) shows the mean values of the energy consumed by the benchmark workloads using the Local, Local\char`_PMCO, and PMCO execution modes. The diagonal brick, diagonal strip, and checkered bars represent the values obtained using PMCO, Local\char`_PMCO, and local execution, respectively.

Figure \ref{fig:secondMulti}(d) depicts the reduction of consumed energy in the PMCO execution mode, which is remarkably smaller than that of the local execution. The execution of the Linpack workload in our experiment consumed 6.24 J using local execution mode and had an overhead of approximately 0.20 J when locally executed using the PMCO. The improvement was substantial, consuming only 2.15 J on the average when executed using PMCO, which is almost 2.5 to 3 times less than the local consumption.

\section{Conclusion}
\label{ch6:con}
In this paper, we proposed the PMCO framework for IoT-supported MCE. Results demonstrated significant time and energy efficiency yield. The performance evaluation of the framework was performed using standard computing benchmarks and custom workloads with eight different intensity levels to highlight the correlation between workload, time, and energy saving effectively. Although the proposed solution was remarkably effective at all intensity levels, the findings became more significant when the workloads were highly intensive. The time and energy efficiency rates increased as the workload intensities were increased.

Results exhibited about 44\% time efficiency and 84\% energy efficiency when the execution of the experimental workload was performed outside the mobile device. The findings were synthesized to demonstrate minimal differences between the reported achievements. The supportive results of real-time experiments revealed the lightweight nature of the framework, as well as its usability and successful adoption in real scenarios. Our secondary experiment confirmed that the lightweight feature of our proposed framework does not deteriorate the performance of the mobile device when used to trigger local execution. In the future, we aim to extend this concept towards mobility assisted edge-to-edge computational offloading mechanism for coping with mobility-related issues posed by IoT devices to save the computational and migration time.

\ifCLASSOPTIONcaptionsoff
  \newpage
\fi

\section*{Acknowledgment}
Imran's work is supported by the Deanship of Scientific Research, King Saud University through Research Group Project number RG-1435-051. This work is funded by Bright Spark Program from the University of Malaya under reference BSP/APP/1635/2013.

\bibliographystyle{IEEEtran}
\bibliography{myrefs.bib}

%\vfill


%\vfill

\begin{IEEEbiographynophoto}{Abdullah Yousafzai}
is an assistant professor with the the Department of Computer Science and Engineering, HITEC University, Taxila, Pakistan. Prior to that, he worked as a Brightspark Research Assistant at C4MCCR University of Malaysia, and as a Backend Web Developer in Pakistan. He received his Ph.D from University of Malaya in 2017, MS (Computer Science) from Comsats Institute of Information Technology, Abbottabad in 2013 and BCS(Hons) from Hazara University Mansehra, Pakistan in 2009. He has reviewed manuscripts for IEEE Access, SUSCOM, TIIS, AJSE, WPC, SUPE and served as a TPC member for numerous international conferences.  His work mainly focuses on Distributed Computing Environments comprising cloud computing systems, edge computing, mobile cloud computing, blockchain systems, and the Internet of Things.
\end{IEEEbiographynophoto}

\begin{IEEEbiographynophoto} {Ibrar Yaqoob} (S'16-M'18-SM'19) is a research professor with the Department of Computer Science and Engineering, Kyung Hee University, South Korea, where he completed his postdoctoral fellowship under the prestigious grant of Brain Korea 21st Century Plus. Prior to that, he received his Ph.D. (Computer Science) from the University of Malaya, Malaysia, in 2017. He worked as a researcher and developer at the Centre for Mobile Cloud Computing Research (C4MCCR), University of Malaya. His numerous research articles are very famous and among the most downloaded in top journals. He has reviewed over 200 times for the top ISI- Indexed journals and conferences. He has been listed among top researchers by Thomson Reuters (Web of Science) based on the number of citations earned in last three years in six categories of Computer Science. He is currently serving/served as a guest/associate editor in various Journals. He has been involved in a number of conferences and workshops in various capacities. His research interests include big data, edge computing, mobile cloud computing, the Internet of Things, and computer networks. 
\end{IEEEbiographynophoto}


\begin{IEEEbiographynophoto} {Muhammad Imran} is working as an Associate Professor in the College of Applied Computer Science, King Saud University (KSU). His research interest includes mobile and wireless networks, Internet of Things, cloud/edge computing, big data analytics, and information security. He has published a number of research papers in refereed international conferences and journals. His research is financially supported by several grants. He served as an Editor in Chief for EAI Transactions on Pervasive Health and Technology. He also serves as an associate editor of many international journals including IEEE Access, IEEE Communications Magazine, and Future Generation Computer Systems. He has been involved in more than seventy-five conferences and workshops in various capacities such as a chair, co-chair and technical program committee member. These include IEEE ICC, Globecom, AINA, LCN, IWCMC, IFIP WWIC and BWCCA. He has received a number of national and international awards.  
\end{IEEEbiographynophoto}

\begin{IEEEbiographynophoto}{Abdullah Gani}
is a full professor at the Department of Computer System and Technology, University of Malaya, Malaysia. His academic qualifications were obtained from the University of Hull, UK for bachelor and master degrees, and the University of Sheffield, UK for Phd. He has vast teaching experience due to having worked in various educational institutions locally and abroad - schools, teaching college, ministry of education, and universities.
His interest in research started in 1983 when he was chosen to attend the Scientific Research Course in RECSAM by the Ministry of Education, Malaysia. More than 200 academic papers have been published in conferences and respectable journals. He actively supervises many students at all level of study - Bachelor, Master and Phd. His interest of research includes self-organized system, reinforcement learning, wireless-related networks. He has worked on mobile cloud computing with High Impact Research Grant of USD 500,000 (RM 1.5M) for the period of 2011-2016. He is a senior member of IEEE.
Currently, he is a director of the Centre for Mobile Cloud Computing Research, which focuses on high impact research. He is also a visiting Professor at the King Saud University, Saudi Arabia as well as serves as Adjunct Professor at the COMSATS Institute of Information Technology, Islamabad, Pakistan. He also serves as a visiting professor at the University Malaysia Sabah, Kota Kinabalu, Sabah. Malaysia (2015-2017). He serves as a chairman of Industry Advisory Panel for Research Degree Program at UNITEN, Malaysia (2015-2017).
\end{IEEEbiographynophoto}

\begin{IEEEbiographynophoto}{Rafidah Md Noor}
received her BIT from University Utara Malaysia in 1998, and M.Sc. in Computer Science from University Technology Malaysia in 2000, and Ph.D. in Computing from Lancaster University in 2010. She is currently a Senior Lecturer at Computer System and Technology Department at the Faculty of Computer Science and Information Technology, University of Malaya. Her research interests include network mobility, vehicular networks, mobile IP, quality of service and quality of experience.
\end{IEEEbiographynophoto}


\end{document}